\documentstyle[epsf,eqsecnum,floats,preprint,aps,epsfig]{revtex}

\input epsf
\tighten
\def\cE{{\cal E}}
\def\cF{{\cal F}}

\begin{document}
\newcommand{\bm}[1]{\mbox{\boldmath{$#1$}}}
\newcommand{\be}{\begin{equation}}
\newcommand{\ee}{\end{equation}}
\newcommand{\bea}{\begin{eqnarray}}
\newcommand{\eea}{\end{eqnarray}}
\newcommand{\barr}{\begin{array}}
\newcommand{\earr}{\end{array}}

\rightline{}
\rightline{hep-th/0406198}
\rightline{UFIFT-HEP-04-10}
\vskip 1cm

\begin{center}
\ \\
\large{{\bf Cosmological Rescaling through Warped Space }} 
\ \\
\ \\
\ \\
\normalsize{ Xingang Chen }
\ \\
\ \\
\small{\em Institute for Fundamental Theory \\ Department of Physics,
University of Florida, Gainesville, FL 32611 }

\end{center}

\begin{abstract}
We discuss a scenario where at least part of the homogeneity on a
brane world can be directly related to the hierarchy problem through
warped space. We study the dynamics of an
anti-D3-brane moving toward the infrared
cut-off of a warped background. After a region described by the DBI
action, the 
self-energy of the anti-D3-brane becomes comparable to the strength of
the
background. Then the world-volume scale of the anti-D3-brane is no
longer 
comoving with the background geometry. After it settles down in the
infrared end, the world-volume inhomogeneity will appear, to a
Poincare observer, to be stretched by
an exponentially large ratio. This ratio is
close to that of the hierarchy problem between the
gravitational and electroweak scales.
\end{abstract}

\setcounter{page}{0}
\thispagestyle{empty}
\maketitle

\eject

\vfill

\baselineskip=18pt

\section{Introduction}
Various hierarchies between different physical scales are among the
most important challenges to any fundamental theory. Two of them
related to this paper are in the context of the particles physics and
cosmology. The former is the hierarchy problem between the
gravitational and electroweak
scale, which is about 32 e-folds. One natural
interpretation to this problem was presented by Randall and Sundrum
using a warped space\cite{Randall:1999ee}. Such a space can be
realized in string
theory\cite{Verlinde:1999fy,Klebanov:2000hb,Giddings:2001yu}
and has also played important roles recently in the studies of the
de Sitter space construction\cite{Kachru:2003aw,Escoda:2003fa}
and brane inflationary
scenario\cite{Dvali:1998pa,Dvali:2001fw,Kachru:2003sx,Hsu:2003cy}.

The latter refers to the hierarchy between the homogeneous
scale of our observable universe and the scale of the big bang
cosmological causal
contact. It is around 30 to 60 e-folds depending on detailed
cosmological evolution. This is usually addressed by
inflationary
scenario\cite{Guth:1980zm}.\footnote{Other attempts
include\cite{Gasperini:1992em,Khoury:2001wf,Chung:1999xg}.} In this
paper, we present a scenario where part of this
hierarchy may be directly related to the one in the particle physics.

We make use of the abovementioned warped space. The string theory
setups in those
studies typically involve some anti-D3-branes located at the infrared
(IR) cutoff of this warped background. 
The ultraviolet (UV) cutoff is smoothly glued into a compact
manifold.
We are living on the
antibranes. Originally these antibranes should enter from the UV
entrance
and be attracted toward the IR end under the background gravitation
and five-form field. Starting with zero initial velocity, the
longitudinal scale of
the antibranes will first comove with the shrinking background and
the inhomogeneities on the antibrane world-volume keep the same
wave-numbers to a Poincare observer. The
proper energy density of the comoving antibranes will grow rapidly and
cause a back-reaction on the background\cite{Silverstein:2003hf}. When
the self-gravitational 
field dominates over the background field, the antibranes enter a
non-comoving phase. Because of this phase, after the antibranes lose
their kinetic energy 
and become at rest at the IR end, the antibrane world-volume embedding
in
the ambient space has been rescaled. From the point of view of the
Poincare
observer, the inhomogeneities on the world-volume have been stretched
by an exponentially big factor, which is directly determined by the
background geometry. This is close to the hierarchy ratio between the
gravitational and electroweak scales. This mechanism, however, does
not apply to the gravitational inhomogeneity in the background
geometry.

\section{Comoving brane scale in the warped space}
\label{SecComoving}
We shall consider the dynamics of a single anti-D3-brane in most part
of the paper. Having a stack of them for our purpose simply changes
the antibrane tension. For simplicity we will neglect the
gravitational coupling and consider only the flat 4-d space-time to
illustrate the idea. This idea can be generalized to the
gravitationally coupled case as well.
An anti-D3-brane in a warped space 
\bea
ds^2 = G_{MN} dX^M dX^N= h^2(r) \left( -dt^2 + d{\bf x}^2
\right) + h^{-2}(r) dr^2 ~, 
~~~h(r)=r/R ~,
\label{WarpedSpace}
\eea
with a four-form potential $C_{\it 4}$ is described by the
Dirac-Born-Infeld (DBI) plus Chern-Simons action\cite{Aharony:1999ti}
\bea
S= T_3 \int d^4 \xi \left[ -\sqrt{ -{\rm det} \left( \partial_a X^M
\partial_b X^N G_{MN} \right) } 
- \frac{1}{4!} \epsilon^{a_1 \cdots a_4}
\partial_{a_1} X^{M_1} \cdots \partial_{a_4} X^{M_4} C_{M_1 \cdots
M_4} \right] ~,
\label{DBICS}
\eea
where $\xi^a$ $(a=0,1,2,3)$ are the world-volume coordinates on the
antibrane, and $X^M$ $(M=0,1,2,3,4)$ are the coordinates of the
ambient ${\rm AdS_5}$ space. The first four components of $X^M$ are
identified with $x^\mu$ $(\mu=0,1,2,3)$ 
and the last $(X^4)$ with $r$ in (\ref{WarpedSpace}). $T_3$ is the
antibrane tension and $R$ is the characteristic length scale of
the AdS space. Here the
non-vanishing 
four-form potential components are $C_{0123} = h^4$.
The $\xi^a$ dependence of $X^M(\xi^a)$ describes how the antibrane is
embedded in the ambient space. We consider here a homogeneous probe
anti-D3-brane 
along the $x^\mu$ directions. The $r$ is then the transverse
position of this probe antibrane. We take the ansatz $X^i=s(t)
\xi^i$ $(i=1,2,3)$ and $X^4=r(t)$ and use the gauge $X^0\equiv t =
\xi^0$. 

The zero and fourth components of the equations of motion corresponding
to the action (\ref{DBICS}) are
\bea
s^3 \frac{d}{dt} \left( \frac{h^6}{\sqrt{h^4-\dot r^2}} \right) +
\frac{d}{dt} \left( s^3 h^4 \right) =0 ~, 
\label{Eom0} \\
\frac{d}{dt} \left( \frac{h^2 \dot r}{\sqrt{h^4-\dot r^2}} \right) +
\frac{2h \partial_r h}{\sqrt{h^4-\dot r^2}} \left( 2h^4-\dot r^2
\right) +
\partial_r (h^4) = 0 ~.
\label{Eom4}
\eea
The other components vanish identically. We note that
Eq.~(\ref{Eom4}) is independent of the antibrane scale factor $s(t)$
and it gives rise to a conserved quantity
\bea
{\cal E} = \frac{h^6}{\sqrt{h^4-\dot r^2}} + h^4 ~. 
\label{EDen}
\eea
Differentiating both sides of Eq.~(\ref{EDen}) and using
Eq.~(\ref{Eom0}), we can see that $s(t) = {\rm const}$. For
convenience we will choose $s(t)=1$ in the rest of the paper. It is
then easy to check that (\ref{EDen}) is in fact the coordinate energy
density in unit of $T_3$, namely $\cE = E/T_3V_3$.

In this paper we will be interested in the Poincare observer. This
observer stays on the anti-D3-brane and uses the space-time
coordinates $(t,{\bf x})$ defined in (\ref{WarpedSpace}). To this
observer, as long as the anti-brane is away from the UV entrance, the
effective four-dimensional Plank mass is approximately constant as in
the Randall-Sundrum model, while the string scale is
red-shifting.\footnote{As we will discuss in
Sec.~\ref{SecNoncomoving}\&\ref{SecModel}, when the back-reaction of
the anti-D3-brane becomes strong and the DBI action fails, this
definition should be understood as replacing the warped factor $h$ in
(\ref{WarpedSpace}) with an effective warped factor felt by the
anti-D3-brane.}

\begin{figure}[t]
\begin{center}
\epsfig{file=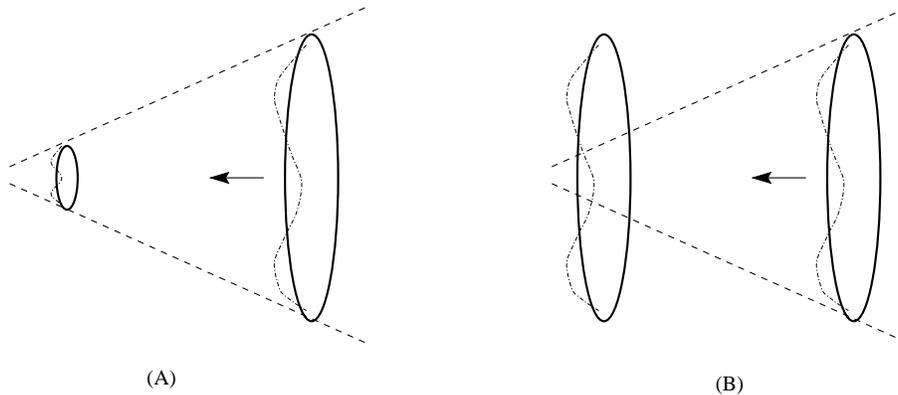, width=12cm}
\end{center} 
\medskip
\caption{The dashed lines are the background geometry. For illustration,
we compactify the three spatial dimensions of the antibrane and show
only one of them on the graph. The wiggles indicate the
inhomogeneities.
In (A), the world-volume scale comoves with the background
geometry, while in (B), the background geometry is negligible to
the antibrane.}
\label{branescale}
\end{figure}

The constant scale factor $s(t)$ means that, according to the DBI
action, the scale of the
antibrane is comoving with the background geometry. (See
Fig.~\ref{branescale}(A).) So if we consider a world-volume
field perturbation,
its wave-number remains the same to a Poincare observer.
This conclusion holds for arbitrary $h(r)$ and $C_{0123}(r)$, and
therefore also applies to D3-branes, as well as other more general
background.
This can also be
straightforwardly 
generalized to the time-dependent background where one replaces the
static
flat metric $\eta_{\mu \nu}$ in (\ref{WarpedSpace}) by $g_{\mu \nu} =
{\rm diag}(-1,a(t),a(t),a(t))$.

The validity of the DBI action requires that the energy involved in
the corresponding low-energy effective field theory
be much smaller than the mass $(\sim r/\alpha')$ of massive W-bosons
that have
been integrated out, namely $\dot r/r \ll
r/\alpha'$\cite{Aharony:1999ti}. It also requires 
that the back-reaction of the probe (anti)branes be negligible 
to the AdS background\cite{Silverstein:2003hf}. This condition will be
important to our discussion.

\section{Comoving region and back-reaction}
\label{BackReaction}
Consider an anti-D3-brane entering the warped space from the UV
entrance ($r=R$) with zero initial velocity (${\cal E}=2$).
As the antibrane travels through the comoving region described by the
DBI action in the last section, it experiences two different phases
before the 
back-reaction becomes large. The first one is the non-relativistic
comoving
phase. This requires $\dot r^2 \ll h^4$ and we can approximate
Eq.~(\ref{EDen}) as 
\bea
\cE \approx \frac{1}{2} \dot r^2 + 2h^4 ~.
\label{NonrelPhase}
\eea
So in this region, $r^4/R^4 \approx 1$ and both constraints mentioned
in Sec.~\ref{SecComoving} can be easily satisfied. 

As $h^4$ decreases, the second term in Eq.~(\ref{EDen}) becomes
negligible when $r^4/R^4 \ll 1$. The antibrane are then entering the
relativistic comoving phase. This 
region is also called the speed limit region and has been discussed in
detail in \cite{Kabat:1999yq,Silverstein:2003hf}. The leading order of
the antibrane motion is completely determined by 
the kinetic term in Eq.~(\ref{EDen}) and this asymptotic behavior is 
\bea
r \rightarrow \frac{R^2}{t} - \frac{R^{10}}{14{\cal E} t^9} + \cdots
~. 
\label{rAsympBeh}
\eea
Using this behavior, the first condition for the validity of the DBI
action mentioned in Sec.~\ref{SecComoving} becomes $R^2 \gg
\alpha'$ , which can be easily met. However the second
requirement, namely the smallness of the back-reaction of this probe
antibrane,
puts a strong constraint on the valid region of the DBI
action\cite{Silverstein:2003hf}.

To a Poincare observer, the
antibrane tension $h^{4}T_3$ is red-shifting as
it travels toward smaller
$r$ region, while from (\ref{EDen}) the total energy density $2T_3$ is
conserved. So at small
$h^4 \ll 1$, this antibrane is highly relativistic. In this process,
the
proper spatial volume of the antibrane shrinks and the energy becomes
more concentrated. If we think of the
AdS background as the near-horizon geometry of $N$ D3-branes ($N$ is
related to $R$ by $R^4 \approx g_s N
\alpha'^2$)\cite{Verlinde:1999fy}, and roughly treat
the gravitational effect of this relativistic antibrane to be similar
to $2h^{-4}$ number of static antibranes, then in order to neglect the
back-reaction we need $h^{4} \gg 2/N$.

For example if $N \approx 10^4$, the relativistic comoving phase is
within 
$10^{-1} \gtrsim r^4/R^4 \gtrsim 10^{-3}$. For $r^4/R^4 \lesssim
10^{-4}$, the 
DBI action starts to break down due to the antibrane
back-reaction. The corresponding warping is approximately
three e-folds. Since the order of the magnitude does not change much
as long as $N$ is not extremely large, we will use this example in the
following. 

For later discussion, we comment that similar behavior applies to a
D3-brane\cite{Silverstein:2003hf}. The difference is that the D3-brane
carries the opposite RR
charge so the Chern-Simons potential terms in Sec.~{\ref{SecComoving}
(such as the second term in Eq.~(\ref{EDen})) change sign. Therefore
D3-brane does not experience a net force in this background. An
initial velocity is necessary for it to fall toward the IR end. For a
small initial velocity $v\ll 1$, the relativistic comoving phase starts
from $h^4 \ll v^2$ and the back-reaction becomes important for $h^4
\lesssim v^2/N$.

\section{Non-comoving antibrane}
\label{SecNoncomoving}
In this section we explore a physical consequence after the antibrane
back-reaction becomes large. We use the AdS background with an IR
cut-off at $r=r_{IR}$\cite{Klebanov:2000hb,Giddings:2001yu}.

If we naively extend the comoving results of Sec.~\ref{SecComoving}
until the warp factor $h$  is several
e-folds below $N^{-1/4}$, the discussion in the last section shows
that the scale of the energy density of the gravitational field of the
probe antibrane is much bigger than that of the
background. Therefore the self-gravitational field of the mobile
antibrane 
dominates over the background and the background fields can be
neglected. Effectively, the DBI action should be modified so that the
induced metric becomes a spatial independent constant, whose value is
roughly determined by the warp factor at the DBI break-down
point. (See Fig.~\ref{branescale}(B).) Realistically there 
should exist a region where the antibrane
transits smoothly from the comoving region to this kind of
behavior. For example in the next section, we will approximate it by a
smooth effective geometry shown in Fig.~\ref{model}. 

Before this relativistic antibrane settles down in the IR end, energy
is lost through the
interactions with the background.
The proper spatial volume ($\int h_e^3 d^3{\bf x}$) of the
antibrane 
should remain approximately the same when it is losing its kinetic
energy in the
non-comoving region where the antibrane self-energy dominates. At the
mean while, the decreasing
antibrane self-energy causes the restoration of the background
geometry. Eventually after the antibrane 
becomes at rest at the IR tip, we can again treat it as a
probe antibrane. But the important thing is that
the embedding of the antibrane in the ambient space 
has now been rescaled. To a Poincare observer, the 
initial inhomogeneities and homogeneous patches on the antibrane are
stretched by an exponentially large rescaling factor. (The scalar
field amplitudes
are kept the same in the non-comoving region since in our normalization
(\ref{WarpedSpace}) they
are proportional to $\Delta r$.)

According to Randall and Sundrum, to the Poincare observer, the
shrinking of the effective background warp 
factor reduces the energy scale of the
electroweak interaction relative to the
gravitational scale. If at the UV entrance we assume that both of the
scales are
of the same magnitude, the final hierarchy is determined by the warp
factor of the IR tip. In our scenario, part of the homogeneity and
flatness of
the universe is achieved in the same cosmological evolution. The
rescaling factor is also completely determined by the background and
is a few e-folds below the hierarchy ratio, since the non-comoving
phase starts several e-folds away from the UV entrance.

We end this section with a few comments.

To have an efficient rescaling of the antibrane world-volume
embedding, it is important that we assume most of the kinetic energy
is lost in the non-comoving region. This is plausible since in this
region the background is distorted by the antibrane and the
interaction between them is maximum. But the details remain to be
understood more explicitly.

This mechanism does not apply to any inhomogeneity existing in the AdS
background, which may give rise to a spatial varying 4-d Planck
constant. To a Poincare observer, this inhomogeneity does not rescale
as the antibrane embedding scale changes.

\section{A simple model}
\label{SecModel}
Comparing to the DBI action, the detailed evolution of the process
described in the last 
section is under less mathematical control so far. Here we give a
simple model to describe a possible intermediate process. We note
that, while some robust features of our scenario have been 
addressed in the last few sections, the more detailed evolution of the
non-comoving phase we are
about to describe is very heuristic and mainly used for illustration.

We assume that the antibrane loses most of its kinetic energy only
when it moves in the non-comoving region around the IR tip which has
the effective warp factor 
$h_e$ due to the antibrane back-reaction. Then it will bounce back (in
terms of the radial coordinate $r$),
and, after a non-comoving phase, enters back to the comoving
phase and reaches a maximum point $h=h_0$. We connect the different
phases smoothly by a straight 
effective geometry as shown in Fig.~\ref{model}. Since $h_e \ll h_0$,
most part of this effective geometry is close to the background
geometry, except for a small region around the IR end, which
effectively describes the non-comoving phase. After the antibrane
reaches $h_0$, it falls back again
due to the background attraction and oscillates in this fashion for
many rounds. The values of $h_e$ and $h_0$ are decreasing
due to the loss of energy. We assume that for each round the fraction
of the energy loss is small enough so we can approximate the evolution
as a smooth function of time.

\begin{figure}[t]
\begin{center}
\epsfig{file=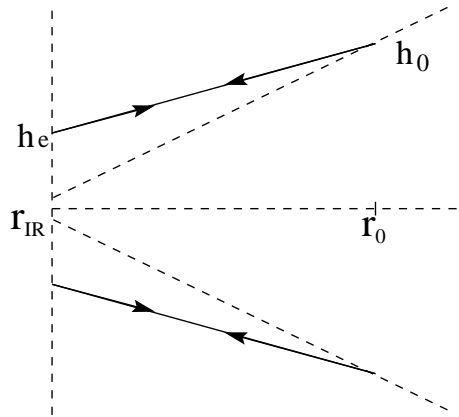, width=6cm}
\end{center} 
\medskip
\caption{The solid lines are the effective geometry taking into
account of the antibrane back-reaction. $r_{IR}$ is the IR cut-off, $r_0$
is the largest $r$-value the antibrane can go. $h_e$ and $h_0$ are the
warp factors of this effective geometry at $r_{IR}$ and $r_0$,
respectively. Both $h_e$ and $h_0$ are decreasing while the
antibrane oscillates.}
\label{model}
\end{figure}

Using (\ref{NonrelPhase}) and (\ref{rAsympBeh}) in our effective
geometry, we see that the time scale of each round is
\bea
\Delta t \approx R h_e^{-1} ~,
\label{Deltat}
\eea
which is dominated by the
speed limit behavior (\ref{rAsympBeh}) at small $r$.

On the other hand, the effective warp factor $h_e$ is determined by
the total antibrane self-energy density around $r_{IR}$. It is
convenient to denote this energy density as
$\cF$ in unit of the static antibrane tension ($\sim h_e^4
T_3$). Since $\cF$ reaches a 
saturation after the antibrane travels several e-folds, $\cF$ cannot
decrease significantly before $h_0$ reaches several e-folds above the
IR tip warp factor $h_{IR}$ of the original background. So for most
range of $h_e$, $\cF$ changes
slowly and we approximate
\bea
\cF \approx \cF_0 + \beta h_e^{\alpha} ~,
\label{cFhe}
\eea
where $\alpha$, $\beta$ are unknown parameters, and $\cF_0$ is the
leading constant. 
If we further assume that the 
fraction of the energy loss for each round is the same,\footnote{If
the 
energy loss has some dependence on $h_e$, it then changes the value of
$\alpha$ in (\ref{Eqnhe}).} from
(\ref{Deltat}) and (\ref{cFhe}) we have
\bea
d h_e \propto - h_e^{-\alpha+2} dt/R ~.
\label{Eqnhe}
\eea
Depending on the value of $\alpha$, $h_e$ decays exponentially or in a
power-law.

Since the proper three-volume of the antibrane remains approximately
unchanged at the IR
tip $r=r_{IR}$, a fixed length scale measured in the world-volume
coordinates 
$\xi^i$ appears to be expanding in the Poincare coordinate $x^i$ as
$h_e$ 
decreases. This is achieved in a purely geometric way.
For example, we label a fixed length on the world-volume in the
1-direction by the world-volume coordinate $\Delta \xi^1$. For the
non-comoving case, the proper length $h_e \Delta x^1$ (for constant
$t$
slice) is approximately fixed during the evolution. We can choose the
convention to have the relation $\Delta \xi^1 = h_e \Delta
x^1$. Therefore for the Poincare observer, the metric can be written
as $ds^2 = -dt^2 + (d{x^i})^2 =
-dt^2 + h_e^{-2} (d{\xi^i})^2$. 
So the effective scale factor $s(t)$ for the
Poincare
observer is proportional to $h_e(t)^{-1}$, which, for world-volume
fields, corresponds to a
universe expanding exponentially or in a power-law. (We note that this
is not inflation since the background metric (\ref{WarpedSpace})
is not inflating.) This is to be
compared with the comoving case where $\Delta x^1=\Delta \xi^1$ as
shown in Sec.~\ref{SecComoving}, in which case, $\Delta x^1$ is fixed
instead of $h_e \Delta x^1$.

\section{Discussion and open questions}
In this paper we have discussed a novel possibility to generate the
brane world-volume homogeneity and flatness. A more detailed
understanding of the intermediate evolution is 
necessary to see how much it can account for the initial conditions
of the big bang. For example, density perturbation is crucial to
possibly connect this mechanism to
observations. Unfortunately we can only make some preliminary
remarks here. In inflation, density fluctuations are seeded when
the scalar field quantum fluctuations exit the horizon and become
classical
objects. Higher frequency modes $k$ exit later and are suppressed, due
to the expansion, by
a factor of $1/k$ relative to the vacuum spectrum of the flat
space. In the rescaling mechanism discussed in this paper,
horizon exit can also happen if the rescaling factor $s(t)$
expands fast enough. If this is the case, the horizon exit of the
density perturbation
and the reheating are happening at the same time -- the background
warp factor $h_e$ restores to $h_{IR}$ at the same time when the
anti-D3-branes
are losing their kinetic energy to string creation. There is also a
suppression for the higher frequency modes -- they exit the horizon
later and therefore correspond to lower kinetic energy. But at the
present stage, it is unclear how much this suppression is in terms of
the density perturbation. 

It is also possible that the rescaling may be combined with the
inflation.
Recent studies indicate that, under certain circumstances, brane
inflation involving D3-anti-D3-branes\cite{Dvali:1998pa,Dvali:2001fw}
can happen in a warped
space\cite{Kachru:2003sx,Hsu:2003cy}. In these models, the
anti-D3-branes have already settled down
in the IR cut-off of the AdS space, while the D3-branes are attracted
toward the IR end from the UV entrance. When
the D3-branes are in their non-relativistic phase, inflation
can happen if the
slow roll conditions can be managed, assuming that the end-product of
the 
brane-anti-brane annihilation leave some anti-D3-branes with a very
small cosmological constant\cite{Kachru:2003aw}. Inflation may also
happen with anti-D3-branes alone at the 
IR end\cite{Pilo:2004mg}. Both the D3 and anti-D3-branes may have
their
non-comoving phase. The rescaling mechanism takes place
before the
inflation for anti-D3-branes, and after the
inflation for D3-branes if they start fast-rolling. This may reduce
the required large inflationary e-foldings. 

Matter (fundamental open strings) can be created on the anti-D3-branes
through various
processes. With the anti-D3-branes alone, we have seen that they gain
kinetic energy from the comoving region. When they interact with the IR
end of the background (or collide with themselves if there are already
some
antibranes in the IR end), presumably part of energy will be used to
create the open strings on the anti-D3-branes. After they settle down,
the open string tension is red-shifted by the corresponding warp
factor, namely $h_{IR}^2 \alpha'$, and the total energy density of
the open 
strings created cannot exceed the kinetic energy density of the
accelerated
anti-D3-brane, which is of order $N h_{IR}^4 T_3$ according to
Sec.~\ref{SecComoving} \& \ref{BackReaction}. If we consider
the abovementioned D3-anti-D3-brane system, D3 branes may gain
similar kinetic energy and transfer part of it to the
anti-D3-branes. In addition, when D3 and anti-D3-branes annihilate,
strings can also be created through the rolling
tachyon\cite{Sen:2002nu}. The tree
level diagram dumps energy into the bulk in terms of the massive
closed strings\cite{Lambert:2003zr}, while loop diagrams connecting
the rolling tachyon and
anti-D3-branes can create massive open strings on the
anti-D3-branes\cite{Chen:2003xq}. The total energy density of the open
strings cannot
exceed the red-shifted (anti)brane tension $n h_{IR}^4 T_3$ where $n$
includes all
the (anti)branes annihilated. Overall we end up
with open strings with red-shifted tension and total energy density at
most of order $N h_{IR}^4 T_3$. This should be responsible for the
subsequent cosmological big bang.

\acknowledgments 
This work was supported in part by the Department of
Energy under Grant No.~DE-FG02-97ER-41029.

\end{document}